\documentclass[pdflatex,sn-mathphys-num]{sn_jnl}

\usepackage[english]{babel}
\usepackage{graphicx}%
\usepackage{multirow}%
\usepackage{amsmath,amssymb,amsfonts}%
\usepackage{amsthm}%
\usepackage{mathrsfs}%
\usepackage[title]{appendix}%
\usepackage{xcolor}%
\usepackage{textcomp}%
\usepackage{manyfoot}%
\usepackage{booktabs}%
\usepackage{algorithm}%
\usepackage{algorithmicx}%
\usepackage{algpseudocode}%
\usepackage{listings}%
\usepackage{xr}%
\usepackage{braket}%
\usepackage[backend=bibtex,style=numeric,sorting=none]{biblatex}  

\makeatletter
\renewcommand{\email}[1]{}
\makeatother

\theoremstyle{thmstyleone}%
\theoremstyle{thmstyletwo}%
\theoremstyle{thmstylethree}%
\usepackage{geometry}
\title{Microkelvin resolution thermometry at the nanometre scale}

\author[1]{\fnm{Jack W.} \sur{Hart}}\email{jwh67@cam.ac.uk}
\author[1]{\fnm{Soham} \sur{Pal}}\email{sp2054@cam.ac.uk}
\author[1]{\fnm{Julien R. E.} \sur{Roth}}\email{jrer3@cam.ac.uk}
\author[1]{\fnm{Katie} \sur{Ninham}}\email{kn429@cam.ac.uk}
\author[1]{\fnm{Abbie H.} \sur{Aleksandrova}}\email{ahd34@cam.ac.uk}
\author[1]{\fnm{Xander} \sur{Peetroons}}\email{xprp2@cam.ac.uk}
\author[2]{\fnm{Soumen} \sur{Mandal}}\email{mandals2@cardiff.ac.uk}
\author[2]{\fnm{Oliver A.} \sur{Williams}}\email{williamso@cardiff.ac.uk}
\author[3]{\fnm{Gavin W.} \sur{Morley}}\email{gavin.morley@warwick.ac.uk}
\author[1]{\fnm{Mete} \sur{Atatüre}}\email{ma424@cam.ac.uk}
\author[1]{\fnm{Helena S.} \sur{Knowles}}\email{hsk35@cam.ac.uk}

\affil[1]{\orgdiv{Cavendish Laboratory}, \orgname{University of Cambridge}, \orgaddress{\street{J. J. Thomson Avenue}, \city{Cambridge}, \postcode{CB3 0HE}, \country{UK}}}
\affil[2]{\orgdiv{Department of Physics and Astronomy}, \orgname{Cardiff University}, \orgaddress{\street{ The Parade}, \city{Cardiff}, \postcode{CF24 3AA}, \country{UK}}}
\affil[3]{\orgdiv{Department of Physics}, \orgname{University of Warwick}, \orgaddress{\street{Gibbet Hill Road}, \city{Coventry}, \postcode{CV4 7AL}, \country{UK}}}

\begin{document}
\maketitle

\begin{abstract}   

Accurate temperature readings of transient events at the nanometer scale are challenging due to the low sensitivity of available sensors. Nanodiamonds containing nitrogen-vacancy (NV) centers have been used for nanoscale thermometry in complex environments, including inside living cells. However, their performance has been limited by short coherence times and low photon counts.
In this work, we use isotopically-purified dual-NV nanodiamonds and a bespoke quantum sensing chip to showcase an order of magnitude improvement in temperature measurement sensitivity compared with previous reports. We demonstrate robust temperature measurements with an error of 682 $\mu$K, experimental sensitivities below 50 mK/$\surd \text{Hz}$ and a shot-noise limited sensitivity of 9.6 mK/$\surd \text{Hz}$. To confirm the utility of these high-performance nanothermometers, we quantify the temperature change induced by the thermometry measurement itself, specifically the optical excitation laser used to probe the NV spin state. In addition, we observe directly at the nanometre scale the transient heating caused by the exothermic mixing of dimethyl sulfoxide in water. Sub-millikelvin resolution and millikelvin sensitivity thermometry unlock the possibility of monitoring minute thermal fluctuations in living systems and assessing catalyst performance at the nanometre scale.

\end{abstract}

\section{Introduction}

Nanoscale sensors are emerging as a transformative technology, particularly in the context of new tools for nanoscale thermometry \cite{zhang2025nanodiamond}. One application is in the life sciences, where temperature is directly coupled to the rate of biochemical reactions and the efficiency of enzymatic proteins \cite{somero2020cellular}. Intracellular temperature gradients, hypothesized to be produced by metabolic activity in mitochondria, remain a source of controversy, as a growing body of experimental evidence \cite{plakhotnik2026nanothermometry} show levels of heating that exceed the expected values extrapolated from theoretical models \cite{baffou2014critique}. 

Nanodiamonds containing negatively-charged nitrogen vacancy defects (NVs) are capable of sensing several parameters on the nanoscale including magnetic \cite{chen2022nanodiamond} and electric field \cite{styles2025all}, pH \cite{sow2020high}, electronic \cite{flinn2026nanodiamond} and nuclear \cite{holzgrafe2020nanoscale} spin species and temperature \cite{simpson2017non}. As nanodiamonds are biologically inert and readily internalized by cells, many of these measurements have been carried out in living systems \cite{van2018nanodiamonds}.

NV thermometry utilizes the temperature-dependent change in the zero-field splitting, $D(T)$, of the NV electronic ground spin states (defined as $m_s = 0$ and $m_s=\pm 1$, of which the latter two are energetically degenerate in the absence of an external magnetic field). Spin state populations are distinguishable due to differences in photoluminescence (PL) following optical excitation caused by a spin-selective non-radiative relaxation pathway (preferentially accessible to the $m_s=\pm1$ spins). In thermometry experiments, a temperature change is inferred by the relative shift in the microwave (MW) frequency that resonantly drives transitions between the spin states. Continuous-wave optically detected magnetic resonance (CW-ODMR) experiments sweep the MW frequency under constant optical excitation and extract the zero-field splitting value as the frequency where the PL shows a resonance. This technique has been successfully employed using ensemble NV nanodiamonds inside living cells, however the sensitivity is limited to $\sim$1 K/$\sqrt{\text{Hz}}$ \cite{sow2025millikelvin} due to broad transition linewidths caused by high strain produced during the manufacturing process and large intrinsic spin-bath noise \cite{barry2020sensitivity}.     

In the presence of an external magnetic field, $B_{ex}$, the energy degeneracy of the $m_s=\pm 1$ states is lifted due to the Zeeman effect, allowing selective driving between the $m_s =0$ and individual $m_s=\pm 1$ states. By leveraging the spin-1 nature of the NV, Toyli \textit{et al.} developed the thermal echo (TE) protocol whereby the phase accumulated by an NV superposition is directly related to the temperature-induced detuning from the transition driving frequencies (see Supplementary Note 2) \cite{toyli2013fluorescence}. In bulk diamond samples, a shot-noise limited sensitivity of 25 mK/$\sqrt{\text{Hz}}$ was inferred from experimental observations, which can be improved upon to 10.1 mK/$\sqrt{\text{Hz}}$ with higher order decoupling sequences \cite{wang2015high}. In nanodiamonds, where surface spin noise, lattice strain, high internal spin impurity concentration and charge state instability all contribute to reduced coherence properties, a similar protocol has been used to demonstrate a sensitivity floor as low as 130 mK/$\sqrt{\text{Hz}}$ \cite{neumann2013high}. Hybridized NV platforms can be used to achieve enhanced temperature sensitivities. By utilizing the change in magnetic field from a nearby static magnetic particle close to its Curie temperature, NVs in diamond nanopillars have reported sensitivities as low as 76 $\mu$K/$\sqrt{\text{Hz}}$ \cite{liu2021ultra}. A similar approach was utilized to reach a sensitivity of 251.5 nK/$\sqrt{\text{Hz}}$ in bulk diamond using a gadolinium magnetic flux concentrator \cite{gong2024hybrid}. However, this methodology could not be translated to a cellular system where the distance between the magnetic particle and the NV is not fixed. All-optical NV-based thermometry, wherein the photon emission around the zero-phonon line is altered by temperature-induced changes to the NV electron-phonon coupling, has demonstrated a sensitivity of 0.3 K/$\sqrt{\text{Hz}}$, but is susceptible to minor changes in background fluorescence \cite{plakhotnik2015all}.

Here, we exploit the long coherence properties of dual-NV nanodiamonds ball-milled from isotopically purified $^{12}$C diamond \cite{march2023long} to perform thermometry protocols at sensitivities equal to and exceeding that typically found in bulk samples \cite{wang2015high}. We demonstrate experimental temperature sensitivities of 48.2 mK/$\sqrt{\text{Hz}}$ under real operating conditions, 9.6 mK/$\surd \text{Hz}$ in the shot noise limit and a resolution floor of 682 $\mu$K. Using an exemplary nanodiamond sensor, we investigate the temperature variations caused by increasing the power of the incident excitation laser and the transient heating caused by the exothermic mixing of dimethyl sulfoxide (DMSO) in water.  

\begin{figure}
\centering
\includegraphics[width=1\linewidth]{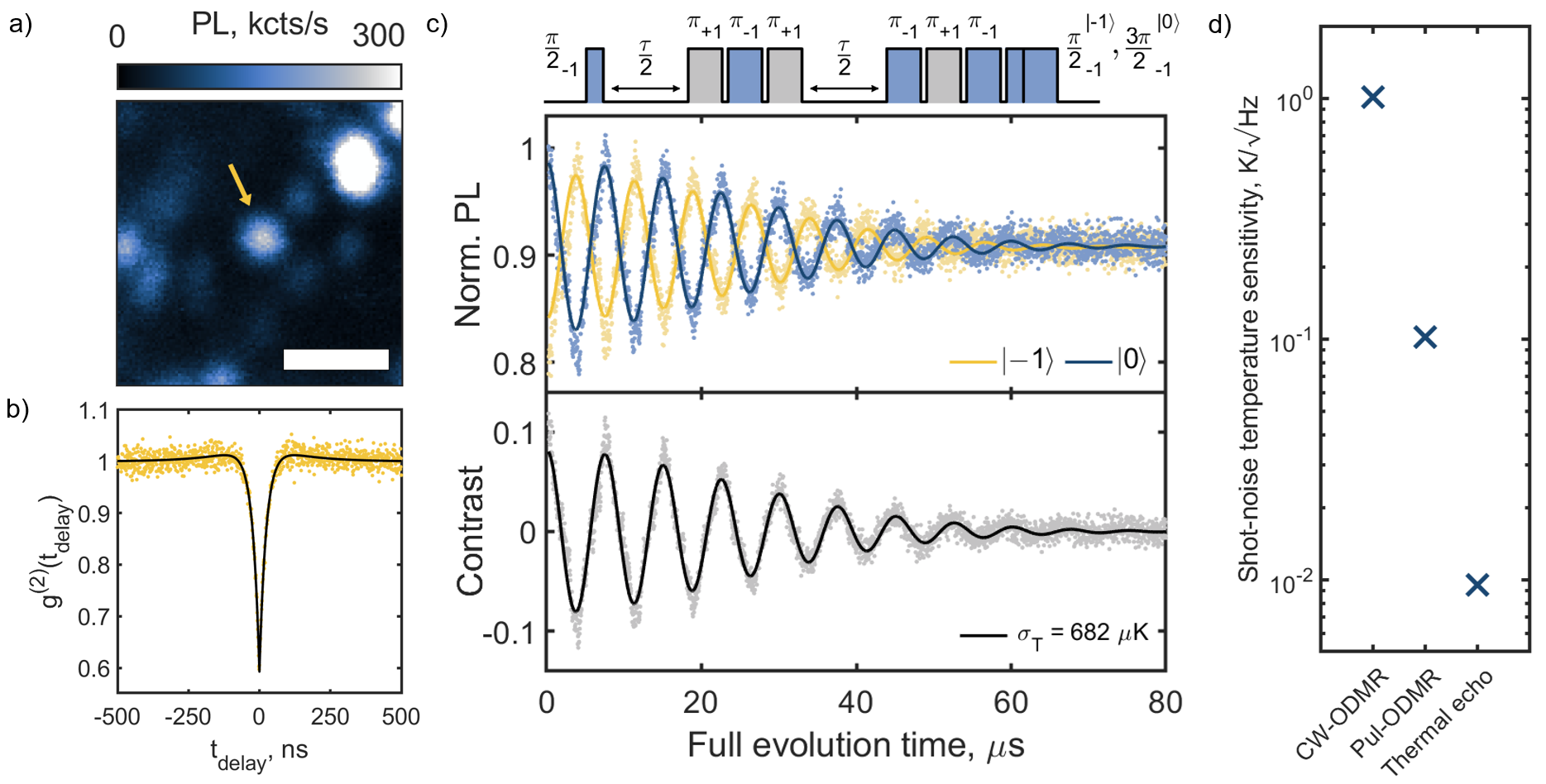}
\caption{Exemplary dual-NV nanodiamond temperature measurements using thermal echo protocol. \textbf{a)} Confocal scan of exemplary nanodiamond fluorescence (highlighted with arrow), scalebar = 2 $\mu m$. \textbf{b)} Normalized correlation between photon detection events as a function of time delay, g$^{(2)}$(t$_{\text{delay}}$), yielding from fit g$^{(2)}$(t$_\text{{delay}}$=0) = 0.592 $\pm$ 0.004 indicating two NV centers within the nanodiamond. 
\textbf{c)} (Top) Thermal echo microwave pulse sequence over a full evolution time, $\tau$, where subscripts correspond to the transition being driven (`-1' for $\ket{0}\rightarrow \ket{-1}$ and `+1' for $\ket{0}\rightarrow \ket{+1}$). The superscripts on the final pulse represent the state projection for readout. (Bottom) Thermal echo photoluminescence (PL) readout for projections in the $\ket{0}$ and $\ket{-1}$ states, with the corresponding contrast below (defined as the difference in state projection PL divided by the sum). The temperature sensing resolution floor is extrapolated from the functional form fitting error (see Equation \ref{eqn:functionalForm}) to be 682 $\mu$K. \textbf{d)} Comparison of the nanodiamond shot-noise limited temperature sensitivities for continuous-wave (CW) ODMR, pulsed (Pul) ODMR and thermal echo sequence respectively (derived from Appendix 1). In comparison to CW-ODMR, the thermal echo protocol offers an improvement in sensitivity exceeding two orders of magnitude.}
\label{fig:Main1}
\end{figure}

\section{Results}
We identify an exemplary nanodiamond, Fig. \ref{fig:Main1}a, on a custom confocal setup (see Supplementary Note 1.1) \cite{binder2017qudi}. The normalized correlation between photon detection events at zero time delay, g$^{(2)}$($t_{delay}$=0) = 0.592 $\pm$ 0.004, indicating the nanodiamond contains two NV centers (Fig. \ref{fig:Main1}b)) \cite{wood2022long}.
To manipulate the population of the NV ground spin states, we apply an external magnetic field of $\sim$5 mT to induce Zeeman splitting of 140 MHz, isolating the two spin transitions of the spin triplet. From the CW-ODMR and magnetic field alignment, we conclude that the two NVs are orientated in the same direction within the diamond lattice. We configure microwave delivery to give Rabi rates between the m$_s=$ 0 and m$_s=\pm$1 spin states of $\sim(2\pi)$10.5 MHz (Fig. S\ref{fig:SuppRabi}). Precise transition frequencies are inferred using a Ramsey protocol, which yields a $T_2^*$ of 11.7 $\pm$ 1.6 $\mu$s (Fig. S\ref{fig:SuppRamsey}). The spin-spin coherence time, extracted using a Hahn-echo protocol, is $T_{2,\text{HE}}$ = 102.6 $\pm$ 0.9 $\mu$s (Fig. S\ref{fig:SuppHahn}). Comparatively the spin-lattice relaxation time, $T_{1}$ = 5.2 $\pm$ 0.3 ms (Fig. S\ref{fig:SuppT1}). Characterization of the nanodiamond size is undertaken by measuring the sample surface topography using an atomic force microscope (Fig. S\ref{fig:SuppAFM}) and corroborated with dynamic light scattering (Fig. S\ref{fig:SuppDLS}) on a dilute sample (see Supplementary Notes 1.2 and 1.3). We determine the nanodiamond to have a major axis diameter of approximately 280 nm, in agreement with the original work \cite{march2023long}.
Nanodiamonds of this size have been reported to be uptaken by cells, therefore retaining their utility despite being larger than other intracellular studies \cite{takahashi2025investigating}. 
\\
Figure \ref{fig:Main1}c (top) shows the TE MW pulse sequence acting on the dual-NV system (where colour and subscript correspond to the transition being driven, i.e. `-1' for $\ket{0}\rightarrow \ket{-1}$ and `+1' for $\ket{0}\rightarrow \ket{+1}$). The corresponding data is taken at an arbitrary low MW detuning to induce oscillations. Phase accumulated during the full evolution time, $\tau$ is projected onto the $\ket{0}$ and $\ket{-1}$ states sequentially during acquisition by altering the length of the final MW pulse. The contrast, defined as the difference between the two PL measurements divided by the sum, removes common mode noise. A temperature noise floor, $\sigma_T$, determined by the error on the frequency component of the functional form fit multiplied by the extracted temperature-frequency proportionality constant (see Appendix), is found to be 682 $\mu$K. Figure \ref{fig:Main1}d compares the shot-noise limited temperature sensitivities we achieve using CW-ODMR, pulsed-ODMR and TE (see Appendix). For the TE protocol, the shot-noise limited sensitivity is 9.6 mK/$\surd \text{Hz}$, comparable with reported bulk diamond values \cite{wang2015high} and an improvement of two orders of magnitude when compared to CW-ODMR. The unique advantage of the dual-NV nanodiamond is double the photon counts with no adverse effects on coherence properties. As such, this nanodiamond gains an approximate 30\% improvement in temperature sensitivity compared with an equivalent single-NV nanodiamond. Whilst this measurement was performed with an aligned external magnetic field, only an $\sim$20\% decrease in temperature resolution and sensitivity is found for a misaligned external field with a polar angle relative to the NV axis of 78° (Fig. S\ref{fig:SuppMisalignedField}). 

\begin{figure}[h]
\centering
\includegraphics[width=1\linewidth]{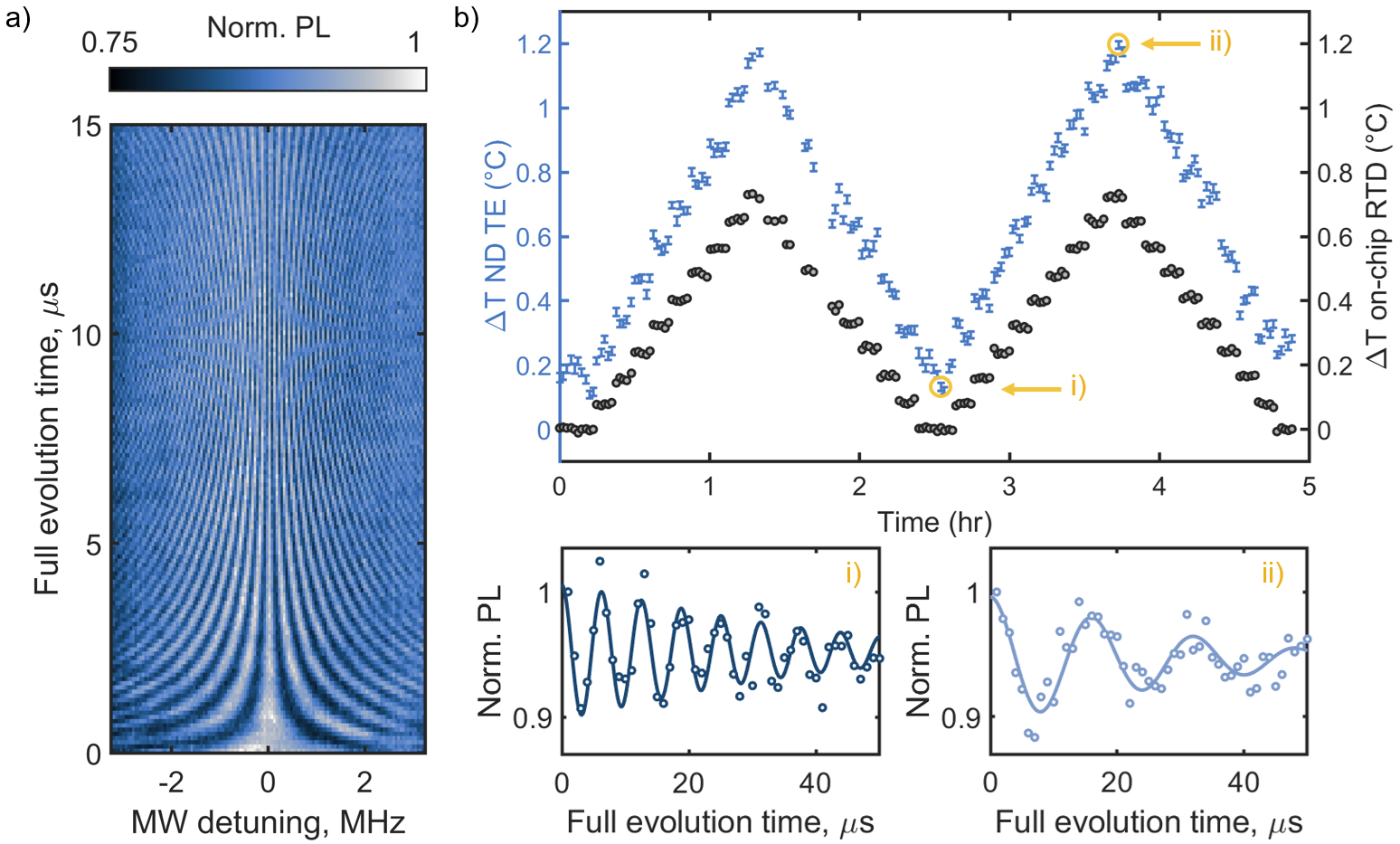}
\caption{\label{fig:Main2}High fidelity temperature sensing using a dual-NV nanodiamond. \textbf{a)} Thermal echo chevron for simulated temperature changes by deliberate microwave (MW) detuning in increments of 50 kHz in a $\pm$3.2 MHz range. \textbf{b)} Temperature ramping using on-chip heating element. Relative changes in the nanodiamond temperature extracted using the thermal echo (blue) closely match the temperature reported by the on-chip resistance temperature detector (RTD, grey circles). Example thermal echo signals at the top (hottest) and bottom (coldest) parts of the heating ramp are show in the subsets i) and ii). }
\end{figure}

\subsection{Robust implementation of nanodiamond thermal echo sensing}
To confirm the robustness and effective sensing bandwidth of the TE protocol on the dual-NV nanodiamond, we deliberately detune the transition MW driving frequencies to simulate a temperature change. Fig. \ref{fig:Main2}a displays the normalized TE signals at detuning increments of 50 kHz in a $\pm$3.2 MHz range (equivalent to temperature difference range of $\sim\pm$33°C). As expected from the theoretical derivation (see Supplementary Note 2), the extracted frequency of the TE signal exactly matches the absolute MW detuning frequency (Fig S\ref{fig:SuppMwDetuning}a). As the detuning frequency increases, the temperature error also increases due to the reduction in PL contrast and undersampling (Fig. S\ref{fig:SuppMwDetuning}b). At low MW detunings, the error spikes due to the period of the oscillation frequency exceeding the coherence envelope of the TE signal, which results in a poor fit. To optimize the measurement errors across the expected temperature range, we carry out sensing experiments with an initial detuning of 200 kHz. 

Over a physiologically relevant range, the change of $D$ with respect to temperature can be modeled as linear \cite{toyli2012measurement}, but the rate of change, $\frac{dD}{dT}$, must be calibrated \textit{in situ} as NVs in nanodiamonds are subject to high strain \cite{plakhotnik2014all} which can lead to significant scaling inhomogeneity \cite{foy2020wide}. By cross-referencing with a calibrated on-chip resistance temperature detector (RTD, see Fig. S\ref{fig:SuppRTD}a) \cite{shanahan2025q}, the proportionality constant is found to be -95.7 kHz/K, which is extracted from the change in TE frequency at different global temperatures caused by an enclosure incubator (see Fig. S\ref{fig:SuppRTD}b). This value for the NV linear temperature coefficients is larger than typically found in  bulk diamond \cite{acosta2010temperature}, but similar to examples reported previously for nanodiamonds \cite{yang2023high,chen2011temperature,foy2020wide}.     

To test the sensing capabilities of the dual-NV nanodiamond, modulations in the local environment temperature are induced by an on-chip heating element. This is implemented using a heating pulse at a fixed power for varying durations between 50 - 500 ns before each TE pulse sequence iteration to achieve a steady state temperature profile during acquisition. The subsequent temperature readout from the nanodiamond, Fig. \ref{fig:Main2}b, shows excellent agreement with the on-chip resistance-temperature detector (RTD), albeit with an offset we attribute to the closer proximity of the nanodiamond to the heating element and the relatively macroscopic size of the RTD (illustrated in Fig. S\ref{fig:SuppWideQBiC}). For this experiment, the transition frequencies are detuned to 200 kHz below resonance, so increases in temperature result in a slower oscillation frequency in the TE signal. The average error on a temperature measurement that took 83 seconds to acquire is 11.9 mK. Exemplary TE data sets from the top (hottest) and bottom (coldest) parts of the heating ramp are shown in the two subsets of Fig. \ref{fig:Main2}b. 

\begin{figure}[h]
\centering
\includegraphics[width=1\linewidth]{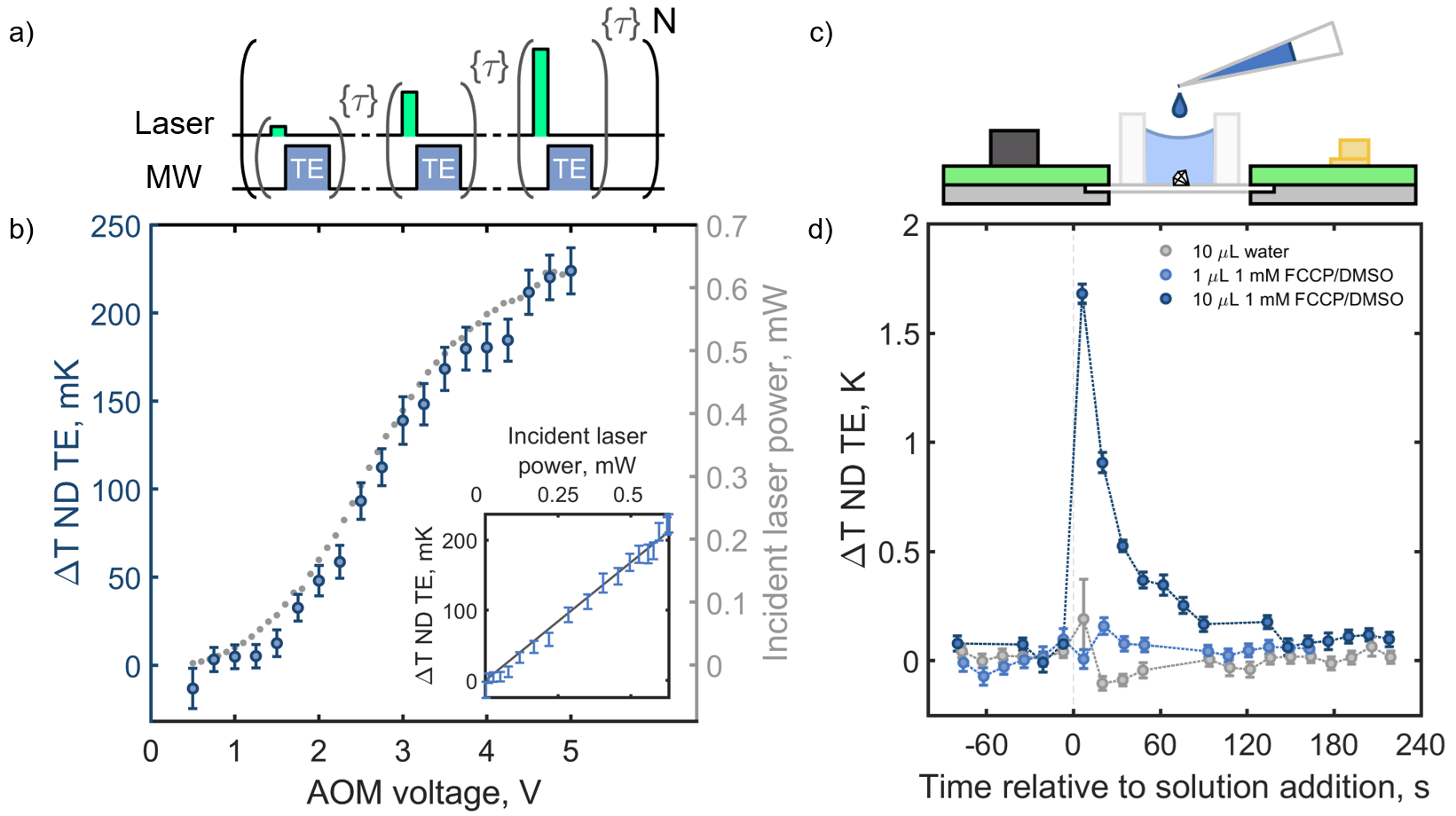}
\caption{\label{fig:Main3}Laser-induced heating and single-shot exothermic chemical temperature sensing experiments. \textbf{a)} Pulse sequence for the laser-induced heating experiment, where the laser power is increased for each set of full evolution times, $\tau$, for N repeats. \textbf{b)} Temperature increase of the nanodiamond as a function of AOM voltage, which corresponds directly to incident laser power. Inset: Linear relationship between incident laser power and temperature at the nanodiamond. \textbf{c)} Schematic of the solution addition experiment. The nanodiamond is submerged in 200 $\mu$L of deionized water in an on-chip well. Solutions are added in such a way as so no contact between the pipette tip and bulk solution is made during addition. \textbf{d)} Temperature change caused by addition of solutions. A significant transient change in temperature is only observed when 10 uL of 1 mM FCCP in DMSO is added to the well. Thermal echo measurements are accumulated every 14 seconds, except in instances where a positional optimization occurred.}
\end{figure}

\subsection{Utility of thermal echo sensing in biologically-relevant experiments}
Unwanted, localized heating is a significant concern when biological samples are exposed to sustained laser illumination such as in super-resolution experiments \cite{waldchen2015light}. To quantify the thermal load applied during quantum sensing experiments, we ramp the power of the 532 nm excitation laser used to initialize and readout the NV fluorescence and measure the temperature reported by the nanodiamond using the TE protocol (Fig. \ref{fig:Main3}a). We fix the laser pulse length at 3 $\mu$s and increased the incident power using the voltage supplied to the acousto-optical modulator (AOM). A corresponding increase in nanodiamond temperature was clearly observed, as shown in Fig. \ref{fig:Main3}b, reaching a maximum temperature rise of 224 mK at 0.63 mW incident laser power (measured at the sample). We observe a linear relationship of 337 mK mW$^{-1}$ between nanodiamond temperature and incident laser power. A proportional increase in nanoparticle temperature with illumination power has been observed in the literature \cite{zograf2017resonant,riviere2022hot}, albeit at significantly higher powers. We attribute the heating observed partially due to the poor thermal conductivity of both the glass substrate the nanodiamond is deposited on and the surrounding ambient air environment \cite{hossain2026optically}.

In experiments probing mitochondrial heating, a chemical uncoupler such as carbonyl cyanide-p-trifluoromethoxyphenylhydrazone (FCCP) is used to induce thermogenesis as it uncouples the oxidative phosphorylation pathway in metabolizing mitochondria \cite{fujiwara2020real}. FCCP is often dissolved in the solvent dimethyl sulfoxide (DMSO), which can induce transient heating when added to an aqueous solution \cite{clever1971enthalpies}. This may be misinterpreted as a mitochondrial thermogenesis event. To quantify this exothermic process and assess its magnitude at the nanoscale, we measure the temperature response of the nanodiamond when submerged in 200 $\mu$L of deionized water during addition of increasing volumes of 1 mM FCCP dissolved in DMSO (Fig.\ref{fig:Main3}c). We observe a significant temperature rise of 1.5 K when a 10 $\mu$L volume of 1 mM FCCP/DMSO solution is added, which then dissipates over 180 seconds (Fig.\ref{fig:Main3}d). 
Whilst the relative magnitude of the heating observed here is similar to reported temperature changes for uncoupled mitochondria, the heating persists for a shorter period of time \cite{fujiwara2020real,lee2025organelle,di2022spatiotemporally}.Different heating phenomena originating from biological activity and chemical perturbation could thus be distinguished using the high temporal resolution and millikelvin resolution of the nanodiamond TE technique.

\section{Conclusion and Outlook}
We report temperature sensing capabilities previously only available in bulk diamond in a nanodiamond system, facilitating sub-millikelvin thermometry with nanometer-scale localization. Utilizing the TE protocol, we extract the linear relationship between optical excitation laser power and nanodiamond temperature. As the resultant heating is low even at saturation laser powers, we infer that our quantum sensing experiments do not produce detrimental heating effects in living systems. Our ability to probe the exothermic mixing of DMSO in water at the nanoscale positions the dual-NV sensor and control chip as a unique platform for assessing the heat production associated with nanoscale chemical and biological catalysts \cite{gao2023emerging,liu2025micro}. This could provide insights into efficacy of a catalytic reaction and a diagnostic tool for designing more efficient catalytic nanostructures \cite{mitchell2021nanoscale}. Furthermore, the robustness to external magnetic field orientation and increased signal-to-noise ratio provided by our platform present an opportunity to investigate mitochondrial thermogenesis at sub-millikelvin resolution in live cells, as temperature perturbations in line with theoretical models are now accessible \cite{macherel2021conundrum}.

\newpage
\section{Appendix 1: Shot noise limit temperature sensitivity calculations}
We define the temperature sensitivity, $\eta$, as
\begin{equation}
    \eta = \delta _T\sqrt{t},
\end{equation} where $\delta_T$ is the minimum measurable temperature difference and $t$ is the corresponding time taken to acquire the measurement. For the experimental results presented in this work, $\delta_ T$ is evaluated from fitting the TE data, $S(\tau)$, to the functional form   
\begin{equation}\label{eqn:functionalForm}
    S(\tau) = C_{TE}.\cos(\omega \tau).\exp\left(-\left(\frac{\tau}{T_{TE}}\right)^n\right),
\end{equation}
where $C_{TE}$ is the signal contrast, $\omega$ is the oscillation angular frequency, $T_{TE}$ is the coherence envelope time constant and $n$ is an envelope stretching term. For experimentally derived sensitivities quoted in the main text, the value of $t$ takes into consideration the computational overhead time associated with loading the pulse sequence as well as the full evolution period, $\tau$.

To calculate the shot-noise limited sensitivity, $\eta_{SN}$, we consider the smallest possible measurable value of temperature, $\delta_T$ as
\begin{equation}
    \delta_T = \delta_S \left|\frac{dS}{dT}\right|^{-1}_{max},
\end{equation}
where $\delta S$ is the shot noise associated with the optical signal acquired and therefore related to the average number of photons collected during a measurement shot, $N_{avg}$, as $\delta S = \frac{1}{\sqrt{N_{avg}}}$. By re-expressing $\frac{dS}{dT} = \frac{dS}{d\omega} \frac{d\omega}{dT}$ and differentiating the functional form for the TE signal in Equation \ref{eqn:functionalForm}, the TE shot noise limited sensitivity, $\eta_{SN,TE}$ can be written as
\begin{equation}\label{eqn:SNsensTE}
    \eta_{SN,TE} = \frac{1}{C_{TE}\sqrt{N_{avg}}} \left|\frac{d\omega}{dT} \right|^{-1} \frac{1}{ \sqrt{\tau} \exp\left(-\left(\frac{\tau}{T_{TE}}\right)^n\right) },
\end{equation}
where for the TE protocol, $\left|\frac{d\omega}{dT}\right| = 2\pi \left|\frac{dD}{dT}\right|$ (see Supplementary Note 2). This can be further simplified by defining the probability of detecting a photon per measurement shot from the NV system when it is in the $\ket{0}$ and $\ket{-1}$ state as $p_0$ and $p_1$ respectively. Now by defining the contrast $C_{TE} =  (p_0-p_1)/(p_0+p_1)$ and the average number of photons $N_{avg}=(p_0+p_1)/2$, the shot-noise-limited TE sensitivity can be expressed as
\begin{equation}
    \eta_{SN,TE} = \sqrt{\frac{2(p_0+p_1)}{(p_0-p_1)^2}} \left| \frac{dD}{dT} \right|^{-1} \frac{1}{2\pi\sqrt{\tau}  \exp\left(-\left(\frac{\tau}{T_{TE}}\right)^n\right)  },
\end{equation}
which is the same result as in \cite{toyli2013fluorescence}. As the nanodiamond contains two NVs with the same orientation, we experimentally measure $p_0$ = 0.151 and $p_1$ = 0.112. As can be seen in Fig. \ref{fig:Main1}c, the measured $T_{TE}$ = 34.4$ \pm $0.2 $\mu$s and using the extrapolated proportionality constant (see Fig. S\ref{fig:SuppRTD}) we find the minimum $\eta_{SN,TE}$ = 9.6 mK/$\sqrt{\text{Hz}}$. 
The TE protocol was also carried out on an ensemble NV bulk diamond sample, but resulted in a significantly worse temperature sensitivity of 266 mK/$\sqrt{\text{Hz}}$ (Fig. S\ref{fig:SuppEnsemble}). A variant on the TE sensing protocol that does not require an external magnetic field is also performed on the exemplary nanodiamond, but also shows poorer performance with a temperature sensitivity of 284 mK/$\sqrt{\text{Hz}}$ (Fig. S\ref{fig:SuppZeroField}).  

To compare thermometry methodologies, we also investigate the shot noise limited sensitivity of ODMR (continuous wave and pulsed) as alternative temperature sensing techniques. Using the same logic as above, we determine the shot-noise-limited sensitivity, $\eta_{SN,ODMR}$, as
\begin{equation}\label{eqn:SNsensPulODMR}
    \eta_{SN,ODMR} = \frac{4}{3\sqrt{3}}\frac{\Delta\upsilon}{C\sqrt{R}}\left|\frac{dD}{dT}\right|^{-1}\sqrt{\frac{t_I+t_\pi+t_r}{t_r}}
\end{equation}
where $\Delta\upsilon$ is the experimentally achieved linewidth of the ODMR transition, $C$ is the ODMR contrast, $R$ is the nanodiamond fluorescence photon count rate per second, $t_I$ is the initialization pulse length, $t_{\pi}$ is the microwave ($\pi$) pulse length, $t_r$ is the readout pulse length, and the numerical prefactor originates from the highest gradient component of the Lorentzian line shape \cite{barry2020sensitivity}. Equation \ref{eqn:SNsensPulODMR} can be applied to continuous-wave ODMR by setting the term that accounts for the readout, microwave drive and initialization duty cycle to 1. Using Equation \ref{eqn:SNsensPulODMR} and the optimal combination of $\Delta\upsilon$ and $C$ we find the minimum sensitivity for CW-ODMR is $\eta_{SN,ODMR}$ = 1.013 K/$\sqrt{\text{Hz}}$, whereas for pulsed-ODMR the minimum sensitivity is $\eta_{SN,ODMR}$ = 102 mK/$\sqrt{\text{Hz}}$ (see Fig. \ref{fig:Main1}d), which is an order of magnitude larger than the TE protocol for the same nanodiamond (Fig. S\ref{fig:SuppPulsedODMR}).

\printbibliography

\end{document}